# The Impact of Regulation Regime Changes on ChiNext IPOs: Effects of 2013 and 2020 Reforms on Pricing and Overreaction

Qi Deng[1,2*], Lunge Dai[3], Zixin Yang[4],
Zhong-guo Zhou[5§], Monica Hussein[5], Dingyi Chen[1†], Mick Swartz[6]

## Abstract

Since its inauguration, ChiNext has gone through three time periods with two different regulation regimes and three different sets of listing day trading restrictions. Till the 2013 reform, ChiNext IPOs are under the "approval" regulation regime with listing day trading curbs; after the 2013 reform, the regulation regime is still of the approval type, but instead a listing day return limit of 44% is imposed on IPOs; after the 2020 reform, ChiNext IPOs are supervised by the "registration" regime with no listing day trading restrictions. This paper studies the impact of regulation regimes and listing day trading restrictions on the initial return of ChiNext IPOs. We hypothesize that the initial return of a ChiNext IPO contains the issuer's intrinsic value and the investors' overreaction. The intrinsic value is represented by the IPO's 21$^{st}$ day return ("monthly return), and the difference between the monthly and initial returns ("intramonth return") is a proxy of the overreaction. We investigate this conjecture and identify variables that are significant for the three returns in the three time periods. We find that all significant variables for all three returns in all three time periods fall into four categories: pre-listing demand, post-listing demand, market condition and pre-listing issuer value. We observe stark contrasts among variable categories for each of the returns in the three time periods, which reveals an evolution of the investors' behavior with regard to the progression of regulation regimes. Based on our findings, we argue that the differences among the levels and determinants of initial return, monthly return (intrinsic value) and intramonth return (overreaction) in different time periods can be largely explained by regulation regime changes along two dimensions: 1) approval vs. registration and 2) listing day trading curbs and return limits. We find that IPO pricing is demand-driven under the approval regime, but value-driven under the registration regime. We further compare the impact of regulation regime changes on ChiNext IPO pricing practice, and propose a future research plan on ChiNext IPO pricing efficiency with policy implication.

†*§ Corresponding author.
1. College of Artificial Intelligence, Hubei University of Automotive Technology, Shiyan, Hubei, China
2. Cofintelligence Financial Technology Ltd., Hong Kong and Shanghai, China
3. Warwick Business School, University of Warwick, Warwick, UK
4. School of Management, University College London, London, UK
5. Department of Finance, Financial Planning, and Insurance, David Nazarian College of Business and Economics, California State University, Northridge, California, USA
6. Marshall School of Business, University of Southern California, Los Angeles, California, USA
† First corresponding author, email: dy.chen@huat.edu.cn
* Second corresponding author, email: dq@huat.edu.cn; qi.deng@cofintelligence.com
§ Third corresponding author, email: zhong-guo.zhou@csun.edu

1. Introduction

China has established a multitier capital market designed for enterprises at different stages of growth and of different quality and risk profiles to raise capital; the system also aims to satisfy different risk appetites of investors. Thus far, China has developed a relatively complete capital market hierarchy, composed of the main boards on Shanghai Securities Exchange (SHSE) and Shenzhen Securities Exchange (SZSE), as well as a variety of second boards that include the Small and Medium Enterprise (SME) and ChiNext boards on SZSE, and the STAR ("Sci-Tech innovAtion boaRd") board on SHSE. The most recent addition to the mix is the Beijing Securities Exchange (BJSE), launched on September 3, 2021, of which the main purpose is to promote liquidity for issuers that have already been listed on the non-market-making, and therefore illiquid "New Third Board."

Our focus is China's original second board that mainly serve the "entrepreneurial" ventures, namely the ChiNext board on SZSE. We choose ChiNext for several reasons. First, ChiNext is China's first second board where a pure demand-driven share subscription and allocation process, known as the "Chinese-style" bookbuilding process, is utilized to price IPOs since its commencement on October 30, 2009. Thus, it is presumably more efficient in IPO pricing than the main boards. Second, the average gap between the IPO pricing date and the listing date for ChiNext is much shorter than that for all other boards (Hussein, Zhou and Deng, 2019). A shorter gap narrows the window for potential market uncertainty and therefore makes IPO pricing less erroneous. Third, aspiring ChiNext issuers are typically young, small, fast-growing and high-tech firms perceived as riskier than their main board counterparts. Therefore, overreaction among ChiNext IPO investors tends to be larger.

The major challenge for an IPO is to effectively determine its offer price: an overpriced IPO dampens demand, while an underpriced IPO undersells the issuer. IPO pricing calls for careful analyses of multiple factors; its complexity has led to extended research aiming to better understand and explain its pricing mechanism, of which the effectiveness is partially measured by the IPO's initial return, or underpricing.

Since its establishment, ChiNext has gone through three time periods with two different regulation regimes and three different sets of listing day trading restrictions. Time period 1 is before the 2013 reform, starting on October 30, 2009, the day ChiNext is launched, till December 12, 2012. In time period 1, the regulation regime governing IPOs is of the "approval" type with listing day trading curb rules. Time period 2 is after the 2013 reform and before the 2020 reform, starting on January 1, 2014 to August 6, 2020, during which the regulation regime is still of the approval type, but with listing day return limit instead of trading curbs. Time period 3 is after the 2020 reform from August 24, 2020 to present, with the "registration" regulation regime, and without listing day trading curbs or return limit.

This paper aims to examine the impact of regulation regimes and listing day trading restrictions on ChiNext IPO initial return and overreaction, with new information and data after the 2013 and 2020 reforms. Deng and Zhou (2016) argue that a ChiNext IPO's initial return is a proxy of overall underpricing, its monthly return reflects a fundamental underpricing, and its intramonth return represents an overreaction. Based on new data and evidence, we modify Deng and Zhou (2016) by recognizing that a ChiNext IPO's initial return consists of two components: the issuer's intrinsic value, represented by the IPO's monthly return, and the overreaction among the investors, represented by its intramonth return (the difference between monthly return and initial return). We then assess the levels of initial return, monthly return and intramonth return, and discover the determinants that drive them, under different regulation regimes. We then compare the levels and drivers of these returns in different regulation regimes in order to understand the impact of regulation regimes on ChiNext IPOs.

We find that all significant variables for all three returns in all three time periods fall into four categories: pre-listing demand, post-listing demand, market condition and pre-listing issuer value. We observe stark contrasts among variable categories for each of the returns in the three time periods, which reveals the evolution of investors' behavior with regard to the progression of regulation regimes. Based on our findings, we argue that the differences among the levels and determinants of initial return, monthly return (intrinsic value) and intramonth return (overreaction) in different time periods can be largely explained by regulation regime changes along two dimensions: 1) approval regime vs. registration regime, and 2) listing day trading curbs and return limits.

In time period 1 (approval regime with listing day trading curbs), IPOs are approved by the regulators and the issuers are perceived to be of "good" quality, thus the investors bid for new shares aggressively without overly examining the issuers' quality. As such, the initial return is largely driven by the pre-listing demand, but the pre-listing issuer value plays a rather small role.

In time period 2 (approval regime with listing day return limits), ChiNext IPOs are still approved by CSRC, thus the investors still bid for new shares aggressively, therefore the pre-listing demand is still significant for the initial return. Because of the listing day return limits, the investors augment their gain through secondary market trading, as such the post-listing demand is also significant.

In time period 3 (registration regime with no listing day trading restrictions), ChiNext IPOs are under the registration regime, that the regulators only check whether the issuers' paperwork is in compliance with the listing rules. All issuer information is self-reported and up to the investors to digest. Therefore, the investors examine the issuers' quality in greater details themselves, and the initial return is determined entirely by the pre-listing issuer value. However, the investors are not professionals in analyzing the self-reported public information, neither do they have sufficient means to acquire private information on the issuers, which cause erroneous pricing.

Our contribution is three-fold. First, we are among the first to study the levels and drivers of ChiNext IPO initial return, intrinsic value and overreaction in the post 2020 reform era, during which the new registration regulation regime has replaced the approval regulation regime, with no listing day trading restrictions. Second, we are among the first to compare the impact of regulation regimes on ChiNext IPOs over the entire lifespan of the board. Third, based on our findings, we propose a future research plan to compare IPO pricing efficiency under different regulation regimes, with potential policy implication on ChiNext IPO practice.

The rest of the paper is organized as follows. Section 2 provides literature review on classic IPO underpricing theories, background on ChiNext IPO regulations, and our research perspective and context. Section 3 covers the dataset and variable selection, with descriptive statistics on the returns. Section 4 provides the statistical models along with empirical results. Section 5 discusses the regulation regimes and their impact on the initial return, intrinsic value and overreaction. Section 6 concludes the paper and further compares the impact of regulation regime changes on ChiNext IPO pricing practice, and discusses future research plan on ChiNext IPO pricing efficiency.

## 2. Literature Review, Regulations, and Research Context

### 2.1 Literature Review on Classic IPO Underpricing Literature

The majority of IPO underpricing literature analyzes the causes of underpricing from an information asymmetry perspective. Berle and Means (1930) propose the principal-agent theory to explain information asymmetry between issuers and underwriters. Baron and Holmström (1980) hold the view that underwriters have a better idea on what the market expects and what investors are willing to pay than issuers, and accordingly lower the offer price to offload their burden of selling new shares. Hu et al. (2021), in a China context, find that underwriters with outstanding reputation can reduce level of underpricing, as they are able to reduce the level of information asymmetry by minimizing the time gap between pricing and listing. Rock (1986) proposes the winner's curse hypothesis, focusing on the effect of information asymmetry between individual investors. Beatty and Ritter (1986) and Welch (1992) argue that uninformed investors demand underpricing to offset the additional risks associated with lack of information.

Bhattacharya (1979) puts forward the signaling theory. Ibbotson (1975) states that "competitive companies will choose leave money on the table." Welch (1989) proposes that an issuer lowers offer price to send a signal to investors that it is of good quality. On the other hand, Michaely and Shaw (1994) reach a complete opposite conclusion.

Some scholars explain the underpricing phenomenon from the perspective of market condition and investor sentiment. Ljungqvist et al. (2006) deem that if a country has an overall health economy, or has released favorable policies, the investor sentiment tends to be optimistic, and issuers take advantage and time their issues accordingly.

A group scholars argue that IPO underpricing also provides means of attracting investors and increasing an issuer's exposure. Demers and Lewellen (2003) find that the degree of IPO underpricing is positively correlated with the attention attracted by an issuer from investors after IPO, which is conducive to the issuer's long-term business success.

**2.2 Regulations Regime Evolution of ChiNext IPOs**

The purpose of regulation is to protect legitimate interests of all participants and maintain the public's confidence, and thus foster healthy development of the market. To that end, the country's regulatory authority, China Securities Regulatory Commission (CSRC) has taken upon itself to improve IPO pricing efficiency with continuous effort over the years. In October 2009, with the establishment of ChiNext, CSRC moved away from a P/E ratio based IPO pricing practice and adopted a "Chinese-style" IPO bookbuilding process, in which the underwriters are allowed to price IPOs, but not allocate shares on their own discretion. For all 355 ChiNext IPOs from ChiNext's commencement to December 31, 2012, Deng and Zhou (2015a) find that the mean initial return is 34.41%, which is far less severe than that in the old P/E ratio regime.

In order to further improve IPO pricing efficiency and to complete another round of overhaul in Chinese stock markets, CSRC halted IPO activities from November 3, 2012 till January 17, 2014. The 438-day suspension is the longest ever for ChiNext[1]. They resumed IPO activities in January 2014, along with a new set of rules[2]. The most relevant rule change that directly affects ChiNext IPOs' initial return is SZSE Bulletin #[2014]54 (June 13, 2014), which specifies that the maximum initial return for an IPO be capped at 44% while the maximum loss be limited at 36%, relative to the offer price, on its listing day. On the other hand, the subsequent day return limits of ±10%, relative to the previous trading day closing price, remain in effect. Zhou, Hussein and Deng (2021) study the 2013 reform and find that the average ChiNext IPO initial return is significantly different before and after the reform (till December 31, 2017), which are 34.50% and around 240%, respectively[3]. This finding reveals that, if anything, the listing day return limits introduced by the 2013 reform derails a relatively efficient initial return distribution into an inefficient one, and the reform actually causes more severe IPO underpricing, pushing ChiNext off its original track to become more market-oriented and efficient.

The regulators may have realized the unintended impact of the 2013 reform. On June 12, 2020, with the release of CSRC Decree #167, CSRC officially starts implementing a full "registration" (as opposed to "approval") process for issuers that seek ChiNext listing[4].

---

[1] The CSRC suspended the IPO market again for another 125 days from July 4, 2015 to November 6, 2015 in order to fine-tune certain practices.

[2] See Zhou, Hussein and Deng (2021) for detailed descriptions of these new rules and regulations. They coin the overhaul the "2013 reform."

[3] After the 2013 reform (and before the 2020 reform), an average ChiNext IPO hits the 44% cap with an extremely low return variation of 1.44% on its listing day. The cumulative return keeps rising at the 10% non-listing day daily maximum steadily over the next 15 to 20 days to reach around 240-245%. Therefore, in the context of this study, for the time period after the 2013 reform and before the 2020 reform (time period 2), the 21$^{st}$ day closing price return (with regard to the offer price) is regarded as the "initial return," and the 42$^{nd}$ day closing price return the "monthly return."

[4] The STAR board launched on June 13, 2019, which listed its first batch of 25 issues on July 22, 2019, is actually the first board in China that adopts a full registration process. The outcome of which was regarded as being encouraging by CSRC, which decided to extend the full registration practice to ChiNext in June 2020.

On the same day, SZSE releases Bulletin #[2020]515, which abolishes the return cap over the first five trading days entirely, and increases the subsequent daily return limit to ±20% ("the 2020 reform")[5]. The 2020 reform provides an opportunity to further advance our contribution to the ChiNext IPO literature, which is the key motivation for this paper.

**2.3 Research Perspective and Context**

This paper expands a particular research stream on ChiNext's IPO pricing and initial return. Deng and Zhou (2015a) study all 352 IPOs from the establishment of ChiNext to December 31, 2012 (time period 1), and find that the initial return is driven by pre-issue multiplier (offline oversubscription rate), issue size (size effect), and stock market condition on the listing day (market momentum). Deng and Zhou (2015b) find that the listing day opening price synthesizes the overall market demand from the institutional and individual investors, and that the market condition prior to listing (market momentum), offer size (size effect), and conditional return variance (asymmetric information) are significant. Deng and Zhou (2016) find that the initial return contains a fundamental underpricing and an overreaction, that the fundamental underpricing is represented by the 21$^{st}$ day return (monthly return), and that the difference between the initial and monthly returns represents the overreaction. Deng and Zhou (2017) examine the current "Chinese-style" bookbuilding process of ChiNext, and establish that an underwriter relies upon institutional investors to discover an issuer's intrinsic value (through a preliminary price inquiry), and that the same underwriter adjusts the preliminary price to establish the final offer price, based on its assessment on the institutional investors' motivations. They further propose a new hybrid of "bookbuilding plus discriminatory-priced auction" approach to improve ChiNext IPO pricing efficiency. Hussein, Zhou and Deng (2019) identify several significant risk factors disclosed in IPO prospectuses that affect the 1$^{st}$ day opening and closing price returns, but not the monthly return, suggesting that these risk factors contribute significantly to the overreaction part, but not the fundamental underpricing portion of the initial return. Zhou, Hussein and Deng (2021) examine ChiNext IPOs' performance before and the 2013 reform (time period 2 with a dataset till Dec 31, 2017); they find that there is even more severe initial underpricing and return volatility after the reform. This research stream provides the perspective and context for this paper.

---
[5] The 2020 reform removes the listing day return cap, thus, the definition of "initial return" is reverted to the closing price return on the listing day after the reform.

We aim to further examine the impact of IPO regulation regime changes on ChiNext IPO initial return, with new information and data after the change of listing day trading restrictions in 2013, and after the IPO regime change from approval-based to registration-based in 2020. We propose that a ChiNext IPO's initial return includes two mutually exclusive components, the issuer's intrinsic value and the investor's overreaction. The intrinsic value is reflected by the IPO's monthly return, and the overreaction is represented by its intramonth return (difference between monthly return and initial return). We seek to assess the levels of initial return, intrinsic value and overreaction, and discover the determinants that drive them, under different regulation regimes. We then compare the impact of different regulation regimes on ChiNext IPO pricing and overreaction.

### 3. Dataset and Descriptive Statistics

**3.1 Data**

We collect ChiNext IPO return and pre-listing firm-level data from Wind Financial database. We establish a total number of 102 potential explanatory variables (independent variables), which are derived from literature (including our own previous work) and regulatory documents, for our initial screen[6]. The dataset includes all pre-issue financials for each of the 1,151 ChiNext IPOs from October 30, 2009 till June 8, 2022. Out of which, 349 samples are between October 30, 2009 and December 31, 2012 (time period 1: pre 2013 reform), 479 samples are from January 1, 2014 to August 6, 2020 (time period 2: post 2013 reform and pre 2020 reform), and 323 samples begin on August 24, 2020 till June 30, 2022 (time period 3: post 2020 reform)[7].

**3.2 Descriptive Statistics**

We provide the descriptive statistics for the initial return, the monthly return, and the intramonth return for time periods 1 to 3, respectively, in Table 1.

In time period 1 (pre 2013 reform), for 349 samples, the mean initial return is 34.08% with a 36.36% standard deviation. The mean monthly return is 29.55% with a 43.47% standard deviation. The intramonth return, is statistically significant with a value of -4.53%. These results essentially duplicate the findings of Deng and Zhou (2016), supporting their argument that if the monthly price represents a short-term equilibrium price, then 4.53% of the initial return may be attributed to the overaction of investors on the listing day.

In time period 2 (post 2013 reform and pre 2020 reform), for 479 samples, the mean initial return is 334.6% with a 233.4% standard deviation[8]. The mean monthly return is

---

[6] To keep this manuscript concise, the full table of all 94 variables is not included. However, it is available upon request, along with all the original and processed data and variables. Interested readers may send email to the corresponding authors for data access.
[7] While the CSRC Decree #167 and SZSE Bulletin #[2020]515 were released on June 12, 2020, they became effective for IPOs listed after August 24, 2022 (inclusive).
[8] Again, after the 2013 reform and before the 2020 reform (time period 2), an 44% return cap is imposed on a ChiNext IPO's listing day return, and its daily maximum gain is limited at the 10% thereafter. A typical ChiNext IPO reaches a less-than-10% daily increase after 15 to 20 trading days. Therefore, in the context of

319% with a 231.1% standard deviation. Comparing to time period 1, it is immediately obvious that both returns are higher by a large margin, which provides strong support to Zhou, Hussein and Deng (2021) with a more complete dataset, that, if anything, the 2013 reform actually makes ChiNext IPO underpricing more severe. The intramonth return is significant but with a relatively small (absolute) value of 15.61%, which also supports Zhou, Hussein and Deng (2021) that a short-term equilibrium seems to occur after one month of trading.

In time period 3 (post 2020 reform), for 323 samples, the mean initial return is 190.6% with a 202.8% standard deviation[9]. The mean monthly return is 138.5% with a 181.6% standard deviation. Both returns are significantly less severe than that in time period 2, but still substantially higher than that in time period 1. These results indicate that the 2020 reform (implementation of IPO registration system and abolishment of listing day return cap) partially achieves the intended goal of suppressing the extremely severe underpricing caused by the 2013 reform. The intramonth return is -52.1%, which has the same sign as the intramonth returns in time periods 1 and 2, but a much larger magnitude, suggesting that the 2022 reform may have actually induced more overaction among investors.

### 4. Variable Selection, Statistical Models and Empirical Results

We first screen a variety of possible independent variables to the utmost extent. The variables cover a wide range of pre-listing financials from IPO prospectus and

---

this study, for this time period, the 21$^{st}$ day closing price return (with regard to the offer price) is defined as the initial return, and the 42$^{nd}$ day closing price return the monthly return.

[9] After the 2020 reform (time period 3), with the listing day return cap being removed, we revert the definition of initial return back to the listing day closing price return, and the definition of monthly return to the 21$^{st}$ day closing price return, same as the pre 2013 reform time period.

firm/investor/underwriter/intermediate information, as well as post-listing trading data. We select these variables based on literature, regulation impact, and novelty. Next, we establish econometric models to investigate determinants of the initial, monthly and intramonth returns in time periods 2 and 3[10]. We then compare results between the three time periods.

**4.1 Variable Selection, Pre-test, and Multivariate Linear Regression Models**

The initial return (*IR*), the monthly return (*MR*), and the intramonth return (*IMR*) of IPO $i$, are defined as follow:

$$IR_i = \left(\frac{1CP_i - OP_i}{OP_i}\right) \times 100\% \qquad (1)$$

$$MR_i = \left(\frac{21CP_i - OP_i}{OP_i}\right) \times 100\% \qquad (2)$$

$$IMR_i = MR_i - IR_i \qquad (3)$$

where $1CP_i$ is IPO $i$'s 1st trading day (listing day) closing price, $21CP_i$ is its 21st trading day closing price, and $OP_i$ is its offer price. All returns are measured in percentage.

We screen a total of 94 possible independent variables[11]. In order to avoid multicollinearity, for any particular group of independent variables with high correlations, we conduct a univariate linear regression on each of the variables to drop those with no statistical significance at 5% level (i.e., *p-value* > 0.05), and for the statistically significant ones (i.e., *p-value* ≤ 0.05), we keep only the variable with the highest adjusted $R^2$ change. We repeat the pre-test procedure for all three dependent variables (*IR*, *MR*, *IMR*) and for all time periods.

---

[10] We do not analyze time period 1 in this paper as it has been thoroughly examined by Deng and Zhou (2015a, 2015b, 2016), to which the results we refer.
[11] We eliminated 15 independent variables with large numbers of missing values. In addition, we eliminated two qualitative variables that cannot be coded for numerical regressions (they are irrelevant anyway). We then conducted pre-test with 85 remaining independent variables.

We then arrange IPOs according to their listing dates and use the following cross-sectional multivariate OLS regressions to analyze the impact from the pre-screened variables on the initial return, monthly return and intramonth return:

$$IR_i = \alpha_i + \sum_{j=1}^{n} \beta_j V_{i,j} + \epsilon_i \qquad (4)$$

where $IR_i$ is the initial return for IPO$_i$, $\alpha_i$ is the regression intercept, $\beta_j$ is the regression coefficient for explanatory variable $V_{i,j}$, and $\epsilon_i$ is the error term.

$$MR_i = \alpha_i + \sum_{j=1}^{n} \beta_j V_{i,j} + \epsilon_i \qquad (5)$$

where $MR_i$ is the monthly return for IPO$_i$, $\alpha_i$ is the regression intercept, $\beta_j$ is the regression coefficient for explanatory variable $V_{i,j}$, and $\epsilon_i$ is the error term.

$$IMR_i = MR_i - IR_i = \alpha_i + \sum_{j=1}^{n} \beta_j V_{i,j} + \epsilon_i \qquad (6)$$

where $IMR_i$ is the intramonth return for IPO$_i$, $\alpha_i$ is the regression intercept, $\beta_j$ is the regression coefficient for explanatory variable $V_{i,j}$, and $\epsilon_i$ is the error term.

**4.2 Determinants of Initial Return**

4.2.1 Time Period 2 (Post 2013 Reform and Pre 2020 reform)

We run multifactor OLS regressions to analyze the impact from the potential variables on the initial return (Equation 4). For time period 2, we identify three significant variables (p-values $\leq$ 0.05): offer price, range of returns (listing to 21$^{st}$ day), and listing day return. The model produces an aggregated adjusted $R^2$ of 0.247 (Table 3 Panel A1).

The variable with the most explanator power, the offer price, has an incremental adjusted $R^2$ of 0.140 and a standardized beta of -0.248. The negative sign indicates that investors are attracted to more "reasonably-priced" issues to maximize returns.

The range of returns reflects the return spread (numerical difference between maximum and minimum daily returns) in the issue's first 21 trading days[12]. It increases the adjusted $R^2$ by 0.074, and has a standardized beta of 0.269, indicating that the initial return is positively related to the return fluctuation in the 21-day span. The range of returns can be regarded as a proxy of secondary market trading activities and therefore post-listing demand on new shares; its being significant suggests that the demand level of a new issue in its first 21-day trading positively affects its initial return.

---
[12] Again, for time period 2 (post 2013 reform and pre 2020 reform), the 21$^{st}$ day closing price return (with regard to the offer price) is used as a proxy for the initial return, thus there is a distribution of 21 daily returns.

The listing day return improves the adjusted $R^2$ by 0.033. It is a proxy of post-listing demand for a new issue: Its standardized beta of 0.207 indicates that the higher the return on the listing day, the higher the market demand for the new shares, thus the higher the initial return. However, the listing day return of time period 2 is almost always 44% due to imposed return limit, thus its explanatory power is rather limited.

The above three significant variables reveal that, in between the 2013 and 2020 reforms, investors are better rewarded by issues with lower offer price and higher post-listing demand. None of the three significant variables is related to pre-listing issuer value (performance and financials) and demand (subscription rates and issue size), as the initial return in time period 2 is the accumulative return of 21 days, during which almost all pre-listing information has been consumed. The impact of market condition is not significant either, as it is completely masked by the very high mean initial return of 334.60%,

4.2.2 Time Period 3 (Post 2020 Reform)

We repeat multivariate OLS regression on the initial return (Equation 4) for time period 3, with 323 ChiNext IPOs using the same set of variables. We identify two significant variables: price-to-book ratio, return on investment two years prior to listing.

The model produces a low aggregated adjusted $R^2$ of 0.082, indicating that there are other significant variables yet to be uncovered (Table 4 Panel A1).

The price-to-book ratio contributes an incremental adjusted $R^2$ of 0.071, with a standardized beta of 0.264. It can be interpreted by the investors as a signal from the underwriters that reflects the issuer's quality. Thus, higher price-to-book ratio translates to a perception of higher projected performance, therefore higher initial return.

The return on investment (ROI) two years prior to listing is the return on accumulative investment over two years immediately before listing. It increases the adjusted $R^2$ by 0.011, with a standardized beta of -0.118, indicating that the initial return is negatively related to the pre-listing 2-year ROI, which seems counterintuitive, as one would expect the opposite. A reasonable interpretation is that investors see higher pre-listing ROI as an indicator of slower post-listing growth potential, as the issuer may adjust to "more sustainable" growth strategy and tactics after IPO, which may negatively impact its short-term performance.

The above two significant variables reveal that, after the 2020 reform, investors are more attentive to issuers' value based on information provided by their prospectus. What is more interesting, however, is that none of the most significant variables in time period 1 (Table 2 Panel A), namely offline subscription ratio and issue size (pre-listing demand), and listing day market (market condition), is significant in time period 3, highlighting a fundamental paradigm shift from a demand-driven IPO pricing to a value-driven one. This may indeed be a consequence of the 2020 reform, which departs from the old approval process to a registration alternative. We will discuss more on this in Section 5.

**4.3 Determinants of Monthly Return (issuer's intrinsic value)**

4.3.1 Time Period 2 (Post 2013 Reform and Pre 2020 reform)

Next we run multivariate linear regressions for the monthly return (Equation 5). For time period 2, the model produces an adjusted $R^2$ of 0.250, and identifies three statistically significant variables: offer price, range of returns (listing to 42$^{nd}$ day), and listing day return. The significant variables for the monthly return are almost the same as those for the initial return (other than the time span of range of returns), with very similar adjusted $R^2$ contributions and standardized betas (Table 3 Panel B1).

This is a bit surprising at first glance, however it confirms the findings of Zhou, Hussein and Deng (2021), that a short-term equilibrium seems to occur after one month of trading, with a more complete dataset for the entire time period 2. The return beyond the first month is thus "stabilized." Therefore, the drivers for the initial return and the monthly return won't be very different, they also provide very similar explanation.

4.3.2 Time Period 3 (Post 2020 Reform)

The regression model for the monthly return in time period 3 produces a very respectable adjusted $R^2$ of 0.843, and identifies three significant variables: max drawdown (of the 1$^{st}$ month), listing day return and price-to-book ratio (Table 4 Panel B1).

The max drawdown in the 1$^{st}$ month of trading after listing contributes an adjusted $R^2$ of 0.740, making it the most significant variable by a large margin. Its standardized beta of 0.857 indicates that it impacts the monthly return in a positive and very visible way. The max drawdown reflects trading volatility, thus level of post-listing demand.

The listing day return is the initial return itself, with an adjusted $R^2$ change of 0.090 and a standardized beta of 0.275. Since it is also a dependent variable in our study, we should not overanalyze it, other than regarding it as a proxy of post-listing demand.

The price-to-book ratio contributes an incremental adjusted $R^2$ of 0.013 and a standardized beta of 0.127. Its impact on the monthly return is similar to that on the initial return, albeit numerically smaller as time progresses. It is perceived by the investors as an indicator of projected performance.

The above three significant variables reveal that, the monthly return is primarily driven by post-listing demand, and to a (much) less degree, pre-listing issuer value. Again, none of the significant variables for the monthly return in time period 1 (Table 2 Panel B) is significant in time period 3. In time period 1, the monthly return is largely driven by pre-listing demand, whereas in time period 3, it is primarily driven by post-listing demand. The pre-listing issuer value has rather minor impact in both time periods. This may also be a result of the 2020 reform of regulation regime from approval to registration. We will discuss more on this in Section 5.

**4.4 Determinants of Intramonth Return (Overreaction)**

4.4.1 Time Period 2 (Post 2013 Reform and Pre 2020 Reform)

The intramonth return is the numerical difference between the monthly return and the initial return. In time period 2, the intramonth return is statistically significant, that there is overreaction in the initial return. We run multivariate linear regressions for the intramonth return (Equation 6). For time period 2, the model produces an adjusted $R^2$ of 0.253, and identifies three statistically significant variables: offer price, range of returns ($21^{st}$ to $42^{nd}$ day), and listing day return. These significant variables are again the same as those for the initial and monthly returns (other than the time span of range of returns), with rather similar adjusted $R^2$ changes and standardized betas (Table 3 Panel C1).

This result, if viewed in conjunction to the descriptive statistics, as well as our explanation on the determinants of the monthly return, may offer interesting insight on the overreaction behavior of investors. The overreaction in time period 2 is small. Although its value of -15.61% is numerically larger than that in time period 1 (-4.53%) in absolute value, it is only a much smaller percentage of the initial return (4.67%) than that in time period 1 (13.30%). This reinforces the notion that, in time period 2, a short-term equilibrium occurs after one month of trading, and that return beyond the first month is rather stabile.

Zhou, Hussein and Deng (2021) show that, in the first month of trading in time period 2 (with partial data), the overreaction among investors builds up almost linearly because of the return limits on listing day and subsequent trading days, until the issue's price reaches a short-term equilibrium. Our study is consistent with their finding and takes one step further. We demonstrate that the "steadily" built-up overreaction in the first month dissipates slowly in the second month. In other words, in time period 2, the initial return is dominated by its intrinsic value element, and overreaction is very minor.

4.4.2 Time Period 3 (Post 2020 Reform)

The intramonth return in time period 3 is statistically significant, indicating there is overreaction in the initial return. The regression model for the intramonth return in time period 3 yields an adjusted $R^2$ of 0.507 and identifies three statistically significant variables: listing day return, listing day turnover, and range of returns (Table 4 Panel C1).

The listing day return (initial return) is the most significant variable with an adjusted $R^2$ contribution of 0.421 and a standardized beta of 0.643. The listing day turnover rate has an adjusted $R^2$ change of 0.046 and a standardized beta of 0.228. Both variables reflect post-listing demand and affect the overreaction positively, that higher return and heavier trading on the listing day mean higher overreaction in the initial return, which makes intuitive sense.

The range of returns increases the adjusted $R^2$ by 0.040, and has a standardized beta of -0.268, indicating that the overreaction is negatively related to return fluctuation of the month. A sensible explanation is that higher overreaction induces higher return and heavier trading activity on listing day, then the overreaction dampens over the first month of trading through a narrower return range (compared to initial return), until it is completely squeezed out and the monthly return converges to a short-term equilibrium (intrinsic value).

The above three significant variables give proof that the overreaction is a result of post-listing demand, and gradually fades away in time period 3. This is very different from time period 1 (Table 2 Panel C), in which the overreaction is primarily driven by market condition and then pre-listing demand. Yet again, we argue that this is an outcome of the 2020 reform that shifts the IPO regulation regime from approval to registration. We will discuss more on this in Section 5.

## 5. Impact of Regulation Regime Changes on ChiNext IPO Returns

We find that all significant variables for all three returns in all three time periods fall into four categories: pre-listing demand, post-listing demand, market condition and pre-listing issuer value. We re-present these variables in Table 5, and together with statistics in Table 1, some clear patterns emerge.

Table 5 shows stark contrast among the three time periods, which reveals the evolution of investors' behavior with regard to the progression of regulation regimes. Based on the findings of Section 4, we argue that the differences among the initial return, the monthly return (intrinsic value) and the intramonth return (overreaction) in different time periods can be largely explained by regulation regime changes along two dimensions: 1) approval vs. registration and 2) listing day trading curbs and return limits.

**5.1 Time Period 1: Approval Regime with Listing Day Trading Curbs**

In time period 1 (pre 2013 reform), the initial return is primarily determined by pre-listing demand (adjusted $R^2$ of 0.360), then market condition (adjusted $R^2$ of 0.130), and modified by pre-listing issuer value in a minor way (adjusted $R^2$ of 0.010) (Table 5 Panel A Initial Return/Time Period 1 columns). In this period, IPOs are "approved" by CSRC, thus, from the investors' viewpoint, the issuers' intrinsic value is "good." Therefore, they bid for new shares aggressively without overly examining the issuers' quality themselves. As such, the initial return is largely driven by the investors' pre-listing demand, with very minor contribution from the pre-listing issuer value. There are intraday trading curbs on the listing day, thus, compared to the other two time periods, the mean initial return is low at 34.08%[13]. The low initial return leaves room for investors to adjust their bidding strategy according to the market condition, which provides complementary information on the likelihood of the new issues being accepted by a broader market base.

---

[13] In time period 1, all ChiNext IPOs are subject to intraday trading suspensions on their listing days. A suspension is either price-triggered, if the trading price is off the opening price by more than 10% either way, or turnover ratio triggered, if the turnover ratio exceeds 50%. The suspension can either be temporary (it is lifted after 30–60 minutes) or permanent (it lasts until the end of the trading day).

Deng and Zhou (2016) demonstrate that the monthly return represents a short-term equilibrium, thus it is the objective measure of the issuers' intrinsic value by the market. In time period 1, the intrinsic value is primarily determined by pre-listing demand (adjusted $R^2$ of 0.320), then market condition (adjusted $R^2$ of 0.140), post-listing demand (adjusted $R^2$ of 0.110) and modified by pre-listing issuer value in a minor way (adjusted $R^2$ of 0.010) (Table 5 Panel B Monthly Return/Time Period 1 columns). The intrinsic value is numerically 86.70% of the initial return. Since the issuers are "approved" as having good quality by CSRC and the investors bid for new shares aggressively, the pre-listing demand provides the most relevant information on their intrinsic value. On the other hand, as the investors have no incentive to thoroughly investigate the issuers under the approval regime, the pre-listing issuer value is only marginal, of which the utility of measuring the issuers' intrinsic value is largely overshadowed by that of the pre-listing demand. The heavy after-market trading for hot issues is reflected through the post-listing demand, which supplements the pre-listing demand by providing the general market's assessment on the issuers' intrinsic value. The mean monthly return of 29.55% in time period 1 is the lowest among all three time periods, allowing the market condition to act as a non-issuer-specific supplemental tool assessing the acceptance level of the new shares.

Deng and Zhou (2016) establish that the intramonth return is a proxy of overreaction in the initial return. More precisely, in the context of our model, it reflects the dissipation of overreaction in the first month of trading. In time period 1, the overreaction counts for 13.30% of the initial return, and its dissipation is primarily determined by market condition (adjusted $R^2$ of 0.130), then pre-listing demand (adjusted $R^2$ of 0.040), pre-listing issuer value (adjusted $R^2$ of 0.020) and post-listing demand (adjusted $R^2$ of 0.010) (Table 5 Panel

C Intramonth Return/Time Period 1 columns). Therefore the overreaction fades away primarily in accordance with the market movement. The information embedded in the pre-listing demand, pre-listing value and post-listing demand have been largely consumed in establishing the intrinsic value, as such their impact on the dissipation of overreaction is dominated by that of the market condition. Therefore, in time period 1, the dissipation of overreaction is mainly a reaction to the market, not the issuer specifics.

**5.2 Time Period 2: Approval Regime with Listing Day Return Limits**

In time period 2 (post 2013 reform and pre 2020 reform), since there are strict return limits on listing day and subsequent trading days, in the context of our study, we use the 21$^{st}$ day closing price return to represent the initial return, which is determined by pre-listing demand (adjusted $R^2$ of 0.140) and post-listing demand (adjusted $R^2$ of 0.107) (Table 5 Panel A Initial Return/Time Period 2 columns). In this period, ChiNext IPOs are still "approved" by CSRC, thus the issuers' quality is still "good" in the eyes of investors. However, because of the listing day return limit of 44%, investors supplement their gain through post-listing trading, thus they are less likely to blindly bid for pre-listing shares, but exercise caution by selecting lower priced issues. The adjusted $R^2$ contribution of pre-listing demand is 0.140, lower than that in time period 1 (adjusted $R^2$ of 0.360), indicates this behavior shift. Since the initial return is really the accumulative return of the first 21 trading days, it is also affected by post-listing demand in that 21-day span. The mean initial return is 334.60%, which is the highest in all three time periods, indicating that the return limits on the listing (44%) and subsequent trading days (10%) greatly fuel investors' desire of acquiring new shares through post-listing trading. That market condition has no impact on initial return can be easily explained by the return limits on listing and subsequent days, which mask any general market related trading pattern. The reason that the pre-listing issuer value plays no role is that the approval system does not incentivize the investors to scrutinize the issuers' quality, much like in time period 1.

In time period 2, we use the 42$^{nd}$ day closing price return as the monthly return (intrinsic value). It is discovered by pre-listing demand (adjusted $R^2$ of 0.131) and post-listing demand (adjusted $R^2$ of 0.119), which are exactly the same drivers for the initial returns (Table 5 Panel B Monthly Return/Time Period 2 columns). This can be easily explained by the fact that the numerical value of the monthly return is 95.33% of that of the initial return, and a short-term equilibrium has been established after one month of trading. A closer look shows that the adjusted $R^2$ of pre-listing demand decreases from 0.140 for the initial return to 0.131 for the monthly return, while that the adjusted $R^2$ of post-listing demand increases from 0.107 to 0.119, which suggests that as time progresses, information embedded in the pre-listing demand gets consumed, while information in the post-listing demand gets built up.

The dissipation of overreaction is decided by pre-listing demand (adjusted $R^2$ of 0.130) and post-listing demand (adjusted $R^2$ of 0.123) (Table 5 Panel C Intramonth Return/Time Period 2 columns). The mean intramonth return in time period 2 is the lowest as a percentage of the initial return at 4.67%, thus overreaction is rather minor. Therefore, in time period 2, a short-term equilibrium occurs after one month of trading, and the dissipation of overreaction afterwards is much like otherwise normal trading driven by demand.

**5.3 Time Period 3: Registration w/o Listing Day Return Limit and Trading Curbs**

In time period 3 (post 2020 reform), the initial return is entirely determined by pre-listing issuer value (adjusted $R^2$ of 0.082) (Table 5 Panel A Initial Return/Time Period 3 columns). Starting from time period 3, ChiNext IPOs are "registered," that CSRC no longer puts a "stamp of approval" on the issuers' quality, it only checks whether their paperwork is in compliance with listing rules, at least on theory. All issuer information is "self-reported" and up to the investors to digest. Thus, from the viewpoint of "uninformed" investors, the issuers' quality is large unknown at the pre-listing stage. Therefore, they are cautious in acquiring pre-listing shares, which explains why pre-listing demand has no impact on the initial return. This is the biggest difference between time period 3 and time period 1 (and for that matter, time period 2). Rather, investors now examine the issuers' quality in greater details themselves. However, investors are not trained in analyzing the self-reported public information on the issuers and do not have sufficient means to acquire private information on the issuers through other channels, they may require to be compensated by higher level of underpricing, which may cause erroneous pricing. This is consistent with Beatty and Ritter (1986) and Welch (1992) in that uninformed investors demand underpricing to offset the risks associated with lack of information. That the monthly return (intrinsic value) counts only 72.67% of the initial return and that the overreaction is very high at 27.33% of the initial return provide evidence to this argument. The mean initial return is 190.60%, which is much higher than that in time period 1 and can be attributed to that there is no longer return limits and trading curbs on the listing day. It may also explain why market condition is not significant, as its effect may be too small and thus masked.

The monthly return counts 72.67% of the overall underpricing and is essentially discovered by post-listing demand (adjusted $R^2$ of 0.830), with a minor adjustment from pre-listing issuer value (adjusted $R^2$ of 0.013) (Table 5 Panel B Monthly Return/Time Period 3 columns). Since investors are uninformed, it is up to the market to assess the issuers' intrinsic value through the post-listing demand on new issues, which produces more accurate pricing. The pre-listing issuer value only plays a minor role in value discovery, as any erroneous information embedded in it gets corrected over the first month of trading. The mean monthly return is 138.50%, which is rather high and masks out any impact from market condition. The dissipation of overreaction is decided entirely by post-listing demand (adjusted $R^2$ of 0.507) (Table 5 Panel C Intramonth Return/Time Period 3 columns).

It is worth mentioning that the initial return is only affected by the pre-listing issuer value and the intrinsic value is essentially determined by the post-listing demand, while the dissipation of overreaction is only shaped by the post-listing demand. This is an expected result in a registration regime with self-reported information disclosure for two reasons, that the information provided by the pre-listing issuer value may be of lower quality than under the approval regime, and that the investors are less sophisticated than the regulators in digesting such information. Therefore, the initial return may contain rather high overreaction, which is subsequently corrected through post-market trading. Essentially it is the market that gauges the pricing of a new issue towards a short-term equilibrium, and squeezes out any bubble along the way. The overreaction is the highest among the three time periods at 27.33% of overall underpricing, a strong indication that the investors are less capable of discovering the issuers' intrinsic value, hence there is sizeable pricing error that manifests as overreaction. The market plays an important role in pricing discovery and error correction.

## 6. Conclusions and Discussions

### 6.1 Conclusions of Findings

This paper investigates the determinants of the initial return, month return, and intramonth return of ChiNext IPOs in three different time periods. Time period 1 is the pre 2013 reform era between October 30, 2009 and December 31, 2012, when the regulation regime governing IPO listing is of the approval type, with intraday trading curbs on the list day. Time period 2 is the post 2013 reform and pre 2020 reform age from January 1, 2014 to August 6, 2020, still under the approval regime with listing day return-limits. Time period 3 is the post 2020 reform time, which begins on August 24, 2020 till present (dataset ends on June 30, 2022).

We find that in time period 1, the mean initial return is 34.08% with a 36.36% standard deviation. The mean monthly initial return is 29.55% with a 43.47% standard deviation. The intramonth return, is statistically significant with a value of -4.53%. These results are consistent with findings of Deng and Zhou (2016). In time period 2, the mean initial return is 334.6% with a 233.4% standard deviation. The mean monthly return is 319% with a 231.1% standard deviation. The intramonth return is significant but with a relatively small value of -15.61%. Our results support Zhou, Hussein and Deng (2021). In time period 3, the mean initial return is 190.6% with a 202.8% standard deviation. The mean monthly initial return is 138.5% with a 181.6% standard deviation. The intramonth return is -52.1%.

We screen a variety of possible independent variables and establish econometric models to uncover determinants of the initial, monthly and intramonth returns in time periods 2 and 3, and refer to Deng and Zhou (2016) for the drivers of the three returns in time period 1. We find that all significant variables for all three returns in all three time periods fall into four categories: pre-listing demand, post-listing demand, market condition and pre-listing issuer value. We then compare results between the three time periods.

We discover stark contrast among the three time periods, which reveals the evolution of investors' behavior with regard to the progression of regulation regimes. We propose that the differences in drivers for the initial return, the monthly return (intrinsic value) and the intramonth return (overreaction) in different time periods can be largely explained by regulation regime changes along two dimensions: 1) approval vs. registration and 2) listing day trading curbs and return limits.

In time period 1 (pre 2013 reform), IPOs are approved by CSRC, thus the investors bid for new shares aggressively without overly examining the issuers' quality themselves. As such, the initial return is largely driven by the pre-listing demand. The pre-listing demand is also most relevant in discovering the intrinsic value, which is supplemented by the post-listing demand that provides the general market's assessment on the intrinsic value. The overreaction dissipates primarily in accordance with the market movement.

In time period 2 (post 2013 reform and pre 2020 reform), there are listing day return limits, therefore we use the 21$^{st}$ day closing price return as the initial return, and the 42$^{nd}$ day closing price return as the monthly return. In this period, ChiNext IPOs are still approved by CSRC, thus the investors still bid for new shares aggressively, therefore the pre-listing demand is significant for the initial return. Because of the listing day return limits, the investors augment their gain through secondary market trading, as such the post-listing demand is also significant. The monthly return is 95.33% of that of the initial return, thus the intrinsic value is also primarily discovered through the pre-listing demand and post-listing demand. The overreaction is the lowest as a percentage of the initial return at 4.67%. The dissipation of overreaction is again decided by pre-listing demand and post-listing demand. Therefore, in time period 2, a short-term equilibrium occurs after one month of trading, and the dissipation of overreaction afterwards is much like otherwise normal trading driven by demand.

In time period 3 (post 2020 reform), ChiNext IPOs are under the registration regime, that the regulators only check whether the issuers' paperwork is in compliance with the listing rules. All issuer information is self-reported and up to the investors to digest. Therefore, the investors examine the issuers' quality in greater details themselves, and the initial return is determined entirely by the pre-listing issuer value. However, the investors are not trained in analyzing the self-reported public information and do not have sufficient means to acquire private information on the issuers, which cause erroneous pricing. That the monthly return (intrinsic value) counts only 72.67% of the initial return and that the overreaction is very high at 27.33% of the initial return provide evidence to this argument. The intrinsic value is essentially discovered by the post-listing demand because the investors are uninformed and it is up to the market to produce mor accurate pricing through post-listing trading. The dissipation of overreaction is decided entirely by the post-listing demand.

**6.2 Discussions of Regulation Regimes and their Impact on ChiNext IPOs**

Based on the above results, in this subsection, we further compare the impact of regulation regimes on ChiNext IPO pricing in general. We then propose a future research approach on ChiNext IPO pricing efficiency. We present the key comparisons in Table 6.

In time period 1, the regulation regime is of the approval type and there are intraday trading curbs on the listing day. At this time, the mandate of the regulators is to foster steady growth of a new capital market that caters to the needs of early-stage, high-tech and high-growth firms, with relatively risker profiles than companies that seek listing on the more traditional boards. Therefore, the emphasis is "investor protection," that CSRC and other regulatory agencies see the investors as amateurs, of which the interest needs to be protected by the authority. Therefore, the regulators not only enforce aspiring issuers to disclose relevant information in carefully prepared prospectuses, but also require them to go through rigorous and painstaking scrutiny before their shares getting listed. Their mission is to eliminate the possibilities of any (perceived) risks at the pre-listing stage. This in turn makes the investors' life easier, they only need to bid for new shares blindly and if they do get shares allocated, they are guaranteed to profit. As such, in time period 1 ChiNext IPO pricing is demand-driven.

However, there are two problems associated with the approval regime. First, CSRC, SZSE and a variety of regulators need to engage a tremendous amount of manpower, resources, and time to complete the IPO listing process, which creates a very deep backlog of want-to-be issuers with a very long waiting time for each of them (Deng and Zhou 2015a), and long wait time produces pricing error. Second and unforeseen at the time, while being thorough in examining the issuers at the pre-listing stage, the regulators do not have sufficient resources or effective tools to keep monitoring them after their shares getting listed. There is no recourse if the investors suffer losses due to the issuers' negligence, mismanagement, or unsatisfactory performance. Neither during this time there is delisting mechanism for listed issuers that no longer meet the listing requirements.

The 2013 reform does not change the regulation regime, therefore in time period 2 the approval regime is still in effect. The pricing practice is also similar. The difference is that the 2013 reform removes the trading curbs, but instead imposes return limits on the listing day (44%). Therefore, time period 2 inherits the built-in problems of the approval regime, and adds its own predicaments, that its pricing process is not very efficient as it takes about two months for the market to establish the intrinsic value of a new issue. For that matter, the 2013 reform is widely regarded as a failure, only to be corrected by another round of market overhaul: the 2020 reform.

The 2020 reform changes the regulation regime from approval to registration, and abolishes the listing day trading curbs and return limits. At this juncture, the mandate of CSRC and other regulatory agencies is to make the IPO market more accessible to capital-thirsty companies that deemed strategically important for the country's technology sector, in a timely fashion. Under the registration regime, the emphasis is "compliance," that issuers are still required to provide comprehensive information on prospectuses (technically by their underwriters), but CSRC and SZSE no longer take the extra effort to verify the behind-the-scene accuracy of these self-reported information. Instead, they only need to make sure their information is in "compliant" with a set of target-designed listing rules in format and content (CSRC Decree #128, CSRC Bulletins #[2020]31-35, SZSE Bulletin #[2020]486, all published on June 12, 2020). The philosophy here is that higher level of compliance induces higher quality of information disclosure, reducing the regulators' workload otherwise required on information verification. In addition, the regulators introduce the concept and practice of "continuous compliance" to ensure that the issuers are in compliance "continuously" after their shares are listed. To that end, CSRC details a set of post-listing rules through CSRC Decree #129 on June 12, 2020. Furthermore, for the first time in ChiNext's history, SZSE puts out delisting rules through SZSE Bulletin #[2020]620 as the last resort to penalize listed issuers that are in an unacceptable level of non-compliance. Also, over the years after the 2020 reform, CSRC and SZSE propel the concept of "self-regulation" or "self-discipline" through a series of rules including SZSE Bulletins #[2022]14 and #[2022]16 issued on January 7, 2022, as well as SZSE Bulletins #[2022]728 and #[2022]729 issued on July 29, 2022. As such, in time period 3, ChiNext IPO pricing is value-driven.

The registration regime seems to have provided perfect solutions to the problems of the approval regime. However, it is not without fault. First, under the registration regime, the issuers are required to produce regulation-compliant IPO documents. However, high-compliance information is not necessarily equivalent to high-quality information. Producing compliance-checked documents may discourage the issuers to disclose information that is unique or cannot be easily fitted into compliant format but otherwise valuable, as volunteering such information may add unnecessary non-compliance risk. Therefore, it is likely that the true quality of the information disclosed in the registration regime is lower than that in the approval regime. Second, the regulators are no longer pressed to examine the regulation-compliant information provided by the issuers and their underwriters, such responsibility now falls on the shoulders of the investors. However, investors are less sophisticated in disseminating these information, therefore there is higher pricing error manifested as large overreaction in the initial return.

While our eventual goal is to answer the question of which regulation regime yields the most efficient IPO pricing, this paper does not provide a sufficient answer. We first need to define what constitutes an efficient IPO pricing. While the initial return of an IPO represents its short-term performance and is a measure of pricing efficiency, the IPO's long-term performance is another measure of pricing efficiency as well. Technically, the initial return reflects pricing inefficiency (level of underpricing), while the long-term performance reflects pricing efficiency, that the more efficient an IPO's pricing is, the higher the shares' long-term performance. This echoes the regulators' desire to ensure that the investors earn a healthy return in the long haul, at the meantime maintain a low volatility market that is less friendly to short-term speculators. Therefore, in order to fully analyze IPO pricing efficiency in ChiNext, we need to study its IPOs' long-term perform in secondary trading as well. We aim to achieve the above goal in a subsequent paper.

**Funding Sources**

The research work conducted by the authors for the purpose of developing this paper is partially funded by Hubei University of Automotive Technology Research Fund No. BK202209.

**Declaration of Interest**

None.

**Table 1 - Descriptive Statistics for Initial Return, Monthly Return and Intramonth Return in Three Time Periods**

This table provides the descriptive statistics for 1151 ChiNext IPOs from Oct 30, 2009 till June 30, 2022. There are 349 IPOs in time period 1 (pre 20013 reform, Oct 30, 2009 to Dec 31, 2012), 479 IPOs in time period 2 (post 2013 reform and pre 2020 reform, Jan 1, 2014 to Aug 6, 2020), and 323 IPOs in time period 3 (post 2020 reform, Aug 24, 2020 to June 30, 2022).

| Time Period | Sample Size | | Mean | Intramonth Return p-Value (2-tailed) | As Percentage of Initial Return | SD | Skewness | Kurtosis | Range | Max | Min |
|---|---|---|---|---|---|---|---|---|---|---|---|
| Time period 1: pre 2013 reform Oct 30, 2009 till Dec 31, 2012 | 349 | initial return (%) | 34.08 | | | 36.36 | 1.37 | 5.41 | 215.60 | 198.90 | -16.68 |
| | | monthly return (%) | 29.55 | | 86.71 | 43.47 | 2.02 | 9.66 | 336.30 | 310.30 | -26.03 |
| | | intramonth return (%) | -4.53 | 0.00 | 13.30 | 24.69 | 2.21 | 19.66 | 273.80 | 211.30 | -62.44 |
| Time period 2: post 2013 reform, pre 2020 reform Jan1, 2014 till Aug 6, 2020 | 479 | initial return (%) | 334.60 | | | 233.40 | 0.94 | 2.87 | 860.90 | 871.30 | 10.36 |
| | | monthly return (%) | 319.00 | | 95.34 | 231.10 | 1.34 | 5.05 | 1320.00 | 1324.00 | 4.04 |
| | | intramonth return (%) | -15.61 | 0.01 | 4.67 | 132.00 | 0.40 | 8.10 | 1188.00 | 675.70 | -512.00 |
| Time period 3: post 2020 reform Aug 24, 2020 till Jun 30, 2022 | 323 | initial return (%) | 190.60 | | | 203.80 | 3.15 | 21.35 | 1967.00 | 1943.00 | -24.62 |
| | | monthly return (%) | 138.50 | | 72.67 | 181.60 | 2.55 | 13.17 | 1398.00 | 1298.00 | -100.00 |
| | | intramonth return (%) | -52.10 | 0.00 | 27.33 | 101.60 | -0.25 | 14.06 | 1268.00 | 569.60 | -698.70 |

# Table 2 - Determinants of Initial Return, Monthly Return and Intramonth Return in Time Period 1

We reconstruct the table from Deng and Zhou (2016) for the initial return, monthly return and Intramonth return in time period 1 (pre 20013 reform, Oct 30, 2009 to Dec 31, 2012), with a sample size of 349 IPOs.

**Panel A – Time Period 1: Initial Return Significant Variables**

| Time Period | Initial Return | | | | |
|---|---|---|---|---|---|
| Time period 1: pre 2013 reform | Non-standardized | | Standardized | p-value | Adjust |
| October 30, 2009 till December 31, 2012 | Beta | S.E. | Beta | | $R^2$ Change |
| (Constant) | 364.150 | 53.010 | | 0.000 | |
| Offline oversubscription | 0.570 | 0.040 | 0.580 | 0.000 | 0.280 |
| Listing day market | 1.600 | 0.190 | 0.310 | 0.000 | 0.130 |
| Issue size | -17.030 | 2.710 | -0.270 | 0.000 | 0.080 |
| Pre-issue P/E ratio | -0.170 | 0.070 | -0.120 | 0.020 | 0.010 |
| Adjusted $R^2$ | | | 0.500 | | |

**Panel B – Time Period 1: Monthly Return Significant Variables**

| Time Period | Monthly Return | | | | |
|---|---|---|---|---|---|
| Time period 1: pre 2013 reform | Non-standardized | | Standardized | p-value | Adjust |
| October 30, 2009 till December 31, 2012 | Beta | S.E. | Beta | | $R^2$ Change |
| (Constant) | 215.900 | 56.450 | | 0.000 | |
| Offline oversubscription | 0.150 | 0.060 | 0.130 | 0.010 | 0.220 |
| Listing day market | 1.330 | 0.230 | 0.220 | 0.000 | 0.120 |
| 1st day trading value | 31.020 | 3.380 | 0.490 | 0.000 | 0.110 |
| Issue size | -39.300 | 3.480 | -0.520 | 0.000 | 0.060 |
| Pricing to listing delay | 1.080 | 0.340 | 0.130 | 0.000 | 0.040 |
| 1st day to 21st day market | 1.150 | 0.220 | 0.190 | 0.000 | 0.020 |
| Number of BOD | -2.020 | 0.990 | -0.070 | 0.040 | 0.010 |
| Adjusted $R^2$ | | | 0.580 | | |

**Panel C – Time Period 1: Intramonth Return Significant Variables**

| Time Period | Intramonth Return | | | | |
|---|---|---|---|---|---|
| Time period 1: pre 2013 reform | Non-standardized | | Standardized | p-value | Adjust |
| October 30, 2009 till December 31, 2012 | Beta | S.E. | Beta | | $R^2$ Change |
| (Constant) | -94.780 | 34.270 | | 0.010 | |
| 1st day to 21st day market | -1.030 | 0.160 | -0.320 | 0.000 | 0.110 |
| Online oversubscription | 0.060 | 0.020 | 0.230 | 0.000 | 0.040 |
| Listing day market | -0.630 | 0.160 | -0.190 | 0.000 | 0.020 |
| 1st day trading value | 4.470 | 1.740 | 0.130 | 0.010 | 0.010 |
| Revenue growth rate previous year | -4.950 | 1.700 | -0.150 | 0.000 | 0.010 |
| Pre-issue profit per share | 7.470 | 3.300 | 0.120 | 0.020 | 0.010 |
| Adjusted $R^2$ | | | 0.200 | | |

## Table 3 - Determinants of Initial Return, Monthly Return and Intramonth Return in Time Period 2

We run multivariate linear regressions (Equations 4, 5, 6) for the initial return, monthly return and intramonth return with the pre-screened significant variables. We report the values of these significant variables in this table, along with their correlation matrices.

**Panel A1 – Time Period 2: Initial Return Significant Variables**

| Time Period | Initial Return | | | | |
|---|---|---|---|---|---|
| Time period 2: post 2013 reform and pre 2020 reform January 1, 2014 till August 6, 2020 | Non-standardized | | Standardized | p-value | Adjust $R^2$ Change |
| | Beta | S.E. | Beta | | |
| (Constant) | 159.886 | 55.375 | | 0.004 | |
| Offer Price | -5.032 | 0.912 | -0.248 | 0.000 | 0.140 |
| Range of Returns - listing to 21st day (%) | 0.058 | 0.009 | 0.269 | 0.000 | 0.074 |
| Listing Day Return (%) | 3.656 | 0.788 | 0.207 | 0.000 | 0.033 |
| Adjusted $R^2$ | | | 0.247 | | |

**Panel A2 – Time Period 2: Significant Variables Correlation Matrix**

| | Offer Price | Range of Returns (%) | Listing Day Return (%) |
|---|---|---|---|
| Offer Price | 1.000 | 0.094 | 0.453 |
| Range of Returns (%) | 0.094 | 1.000 | -0.043 |
| Listing Day Return (%) | 0.453 | -0.043 | 1.000 |

**Panel B1 – Time Period 2: Monthly Return Significant Variables**

| Time Period | Monthly Return | | | | |
|---|---|---|---|---|---|
| Time period 2: post 2013 reform and pre 2020 reform January 1, 2014 till August 6, 2020 | Non-standardized | | Standardized | p-value | Adjust $R^2$ Change |
| | Beta | S.E. | Beta | | |
| (Constant) | 140.719 | 54.703 | | 0.010 | |
| Offer Price | -4.699 | 0.901 | -0.234 | 0.000 | 0.131 |
| Range of Returns - listing to 42nd day (%) | 0.063 | 0.009 | 0.295 | 0.000 | 0.088 |
| Listing Day Return (%) | 3.530 | 0.779 | 0.202 | 0.000 | 0.031 |
| Adjusted $R^2$ | | | 0.250 | | |

**Panel B2 – Time Period 2: Significant Variables Correlation Matrix**

| | Offer Price | Range of Returns (%) | Listing Day Return (%) |
|---|---|---|---|
| Offer Price | 1.000 | 0.094 | 0.453 |
| Range of Returns (%) | 0.094 | 1.000 | -0.043 |
| Listing Day Return (%) | 0.453 | -0.043 | 1.000 |

**Panel C1 – Time Period 2: Intramonth Return Significant Variables**

| Time Period | Intramonth Return | | | | |
|---|---|---|---|---|---|
| Time period 2: post 2013 reform and pre 2020 reform January 1, 2014 till August 6, 2020 | Non-standardized | | Standardized | p-value | Adjust $R^2$ Change |
| | Beta | S.E. | Beta | | |
| (Constant) | 84.500 | 54.706 | | 0.123 | |
| Offer Price | -4.590 | 0.901 | -0.228 | 0.000 | 0.130 |
| Range of Returns - 21st to 42nd day (%) | 0.063 | 0.009 | 0.295 | 0.000 | 0.089 |
| Listing Day Return (%) | 3.702 | 0.779 | 0.212 | 0.000 | 0.034 |
| Adjusted $R^2$ | | | 0.253 | | |

**Panel C2 – Time Period 2: Significant Variables Correlation Matrix**

| | Offer Price | Range of Returns (%) | Listing Day Return (%) |
|---|---|---|---|
| Offer Price | 1.000 | 0.094 | 0.453 |
| Range of Returns (%) | 0.094 | 1.000 | -0.043 |
| Listing Day Return (%) | 0.453 | -0.043 | 1.000 |

## Table 4 - Determinants of Initial Return, Monthly Return and Intramonth Return in Time Period 3

We run multivariate linear regressions (Equations 4, 5, 6) for the initial return, monthly return and intramonth return with the pre-screened significant variables. We report the values of these significant variables in this table, along with their correlation matrices.

**Panel A1 – Time Period 3: Initial Return Significant Variables**

| Time Period | Initial Return | | | | |
|---|---|---|---|---|---|
| Time period 3: post 2020 reform August 24, 2020 till June 30, 2022 | Non-standardized | | Standardized | p-value | Adjusted $R^2$ Change |
| | Beta | S.E. | Beta | | |
| (Constant) | 158.919 | 29.310 | | 0.000 | |
| Price to Book Ratio | 21.325 | 4.314 | 0.264 | 0.000 | 0.071 |
| Return on Investment two years prior to listing (%) | -1.994 | 0.904 | -0.118 | 0.028 | 0.011 |
| Adjusted $R^2$ | | | 0.082 | | |

**Panel A2 – Time Period 3: Significant Variables Correlation Matrix**

| | Price to Book Ratio | Return on Investment two years prior to listing (%) |
|---|---|---|
| Price to Book Ratio | 1.000 | 0.068 |
| Return on Investment two years prior to listing (%) | 0.068 | 1.000 |

**Panel B1 – Time Period 3: Monthly Return Significant Variables**

| Time Period | Monthly Return | | | | |
|---|---|---|---|---|---|
| Time period 3: post 2020 reform August 24, 2020 till June 30, 2022 | Non-standardized | | Standardized | p-value | Adjusted $R^2$ Change |
| | Beta | S.E. | Beta | | |
| (Constant) | -22.186 | 8.293 | | 0.008 | |
| Max Drawdown | 0.908 | 0.025 | 0.857 | 0.000 | 0.740 |
| Listing Day Return (%) | 2.709 | 0.226 | 0.275 | 0.000 | 0.090 |
| Price to Book Ratio | 9.195 | 1.717 | 0.127 | 0.000 | 0.013 |
| Adjusted $R^2$ | | | 0.843 | | |

**Panel B2 – Time Period 3: Significant Variables Correlation Matrix**

| | Max Drawdown | Listing Day Return (%) | Price to Book Ratio |
|---|---|---|---|
| Max Drawdown | 1.000 | 0.182 | -0.319 |
| Listing Day Return (%) | 0.182 | 1.000 | -0.219 |
| Price to Book Ratio | -0.319 | -0.219 | 1.000 |

**Panel C1 – Time Period 3: Intramonth Return Significant Variables**

| Time Period | Intramonth Return | | | | |
|---|---|---|---|---|---|
| Time period 3: post 2020 reform August 24, 2020 till June 30, 2022 | Non-standardized | | Standardized | p-value | Adjusted $R^2$ Change |
| | Beta | S.E. | Beta | | |
| (Constant) | 3.634 | 7.184 | | 0.613 | |
| Listing Day Return (%) | 3.546 | 0.220 | 0.643 | 0.000 | 0.421 |
| Listing Day Turnover Rate (%) | 0.776 | 0.141 | 0.228 | 0.000 | 0.046 |
| Range of Returns (%) | -0.111 | 0.017 | -0.268 | 0.000 | 0.040 |
| Adjusted $R^2$ | | | 0.507 | | |

**Panel C2 – Time Period 3: Significant Variables Correlation Matrix**

| | Listing Day Return (%) | Listing Day Turnover Rate (%) | Range of Returns (%) |
|---|---|---|---|
| Listing Day Return (%) | 1.000 | -0.117 | -0.054 |
| Listing Day Turnover Rate (%) | -0.117 | 1.000 | -0.278 |
| Range of Returns (%) | -0.054 | -0.278 | 1.000 |

# Table 5 – Comparisons of Determinant Categories of Initial Return, Monthly Return and Intramonth Return in Three Time Periods

We compare the categories of significant variables for the initial return, monthly return and intramonth return across three time periods. We report the group values of these categories in this table.

### Panel A - Initial Return: Comparisons of Determinant Categories in Three Time Periods

| Initial Return: Mean: 34.08%, SD: 36.36% Model Fit and Significant Variables | Time Period 1 Standardized Beta | Adjust R² Change | Group Adjust R² | Initial Return: Mean: 334.60%, SD: 233.40% Model Fit and Significant Variables | Time Period 2 Standardized Beta | Adjust R² Change | Group Adjust R² | Initial Return: Mean: 190.60%, SD: 203.80% Model Fit and Significant Variables | Time Period 3 Standardized Beta | Adjusted R² Change | Group Adjust R² |
|---|---|---|---|---|---|---|---|---|---|---|---|
| **Pre-listing Demand** | | | | **Pre-listing Demand** | | | | | | | |
| Offline oversubscription | 0.580 | 0.280 | | Offer Price | -0.248 | 0.140 | 0.140 | | | | |
| Issue size | -0.270 | 0.080 | 0.360 | | | | | | | | |
| | | | | **Post-listing Demand** | | | | | | | |
| | | | | Range of Returns - listing to 21st day (%) | 0.269 | 0.074 | 0.107 | | | | |
| | | | | Listing Day Return (%) | 0.207 | 0.033 | | | | | |
| **Market Condition** | | | | | | | | | | | |
| Listing day market | 0.310 | 0.130 | 0.130 | | | | | | | | |
| **Pre-listing Issuer Value** | | | | | | | | **Pre-listing Issuer Value** | | | |
| Pre-issue P/E ratio | -0.120 | 0.010 | 0.010 | | | | | Price to Book Ratio | 0.264 | 0.071 | |
| | | | | | | | | Return on Investment two years prior to listing (%) | -0.118 | 0.011 | 0.082 |
| Adjusted R² | | 0.500 | | Adjusted R² | | 0.247 | | Adjusted R² | | 0.082 | |

### Panel B - Monthly Return: Comparisons of Determinant Categories in Three Time Periods

| Monthly Return Mean: 29.55%, SD: 43.47% Model Fit and Significant Variables | Time Period 1 Standardized Beta | Adjust R² Change | Group Adjust R² | Monthly Return Mean: 319.00%, SD: 231.10% Model Fit and Significant Variables | Time Period 2 Standardized Beta | Adjust R² Change | Group Adjust R² | Monthly Return Mean: 138.50%, SD: 181.60% Model Fit and Significant Variables | Time Period 3 Standardized Beta | Adjust R² Change | Group Adjust R² |
|---|---|---|---|---|---|---|---|---|---|---|---|
| **Pre-listing Demand** | | | | **Pre-listing Demand** | | | | | | | |
| Offline oversubscription | 0.130 | 0.220 | | Offer Price | -0.234 | 0.131 | 0.131 | | | | |
| Issue size | -0.520 | 0.060 | 0.320 | | | | | | | | |
| Pricing to listing delay | 0.130 | 0.040 | | | | | | | | | |
| **Post-listing Demand** | | | | **Post-listing Demand** | | | | **Post-listing Demand** | | | |
| Listing day trading value | 0.490 | 0.110 | 0.110 | Range of Returns - listing to 42nd day (%) | 0.295 | 0.088 | 0.119 | Max Drawdown | 0.857 | 0.740 | 0.830 |
| | | | | Listing Day Return (%) | 0.202 | 0.031 | | Listing Day Return (%) | 0.275 | 0.090 | |
| **Market Condition** | | | | | | | | | | | |
| Listing day market | 0.220 | 0.120 | | | | | | | | | |
| Listing day to 21st day market | 0.190 | 0.020 | 0.140 | | | | | | | | |
| **Pre-listing Issuer Value** | | | | | | | | **Pre-listing Issuer Value** | | | |
| Number of BOD | -0.070 | 0.010 | 0.010 | | | | | Price to Book Ratio | 0.127 | 0.013 | 0.013 |
| Adjusted R² | | 0.580 | | Adjusted R² | | 0.250 | | Adjusted R² | | 0.843 | |

# Table 5 – Comparisons of Determinant Categories of Initial Return, Monthly Return and Intramonth Return in Three Time Periods – con'd

We compare the categories of significant variables for the initial return, monthly return and intramonth return across three time periods. We report the group values of these categories in this table.

**Panel C - Intramonth Return: Comparisons of Determinant Categories in Three Time Periods**

| Intramonthly Return Mean: -4.53%, SD: 24.69% Model Fit and Significant Variables | Time Period 1 Standardized Beta | Adjust R² Change | Group Adjust R² | Intramonthly Return Mean: -15.61%, SD: 132.00% Model Fit and Significant Variables | Time Period 2 Standardized Beta | Adjust R² Change | Group Adjust R² | Intramonthly Return Mean: -52.10%, SD: 101.60% Model Fit and Significant Variables | Time Period 3 Standardized Beta | Adjusted R² Change | Group Adjust R² |
|---|---|---|---|---|---|---|---|---|---|---|---|
| Pre-listing Demand | | | | Pre-listing Demand | | | | | | | |
| Online oversubscription | 0.230 | 0.040 | 0.040 | Offer Price | -0.228 | 0.130 | 0.130 | | | | |
| Post-listing Demand | | | | Post-listing Demand | | | | Post-listing Demand | | | |
| Listing day trading value | 0.130 | 0.010 | 0.010 | Range of Returns - 21st to 42nd day (%) | 0.295 | 0.089 | 0.123 | Listing Day Return (%) | 0.643 | 0.421 | 0.507 |
| | | | | Listing Day Return (%) | 0.212 | 0.034 | | Listing Day Turnover Rate (%) | 0.228 | 0.046 | |
| | | | | | | | | Range of Returns (%) | -0.268 | 0.040 | |
| Market Condition | | | | | | | | | | | |
| Listing day to 21st day market | -0.320 | 0.110 | 0.130 | | | | | | | | |
| Listing day market | -0.190 | 0.020 | | | | | | | | | |
| Pre-listing Issuer Value | | | | | | | | | | | |
| Revenue growth rate previous year | -0.150 | 0.010 | 0.020 | | | | | | | | |
| Pre-issue profit per share | 0.120 | 0.010 | | | | | | | | | |
| Adjusted R² | | 0.200 | | Adjusted R² | | 0.253 | | Adjusted R² | | 0.507 | |

## Table 6 – Comparisons of Regulation Regimes and Their Impact on ChiNext IPOs

We compare the categories of significant variables for the initial return, monthly return and intramonth return across three time periods. We report the group values of these categories in this table.

| Time Period<br>Regulation Regim | Time Period 1<br>Pre 2013 Reform<br>Oct 30, 2009 – Dec 31, 2012 | Time Period 2<br>Post 2013 Reform, Pre 2020 Reform<br>Jan 1, 2014 – Aug 6, 2020 | Time Period 3<br>Post 2020 Reform<br>Aug 24, 2020 – Jun 30, 2022 |
|---|---|---|---|
| Regulation regime | Approval | Approval | Registration |
| Listing day trading restrictions | Intraday tradig curbs | Return limit of 44% | None |
| Regulator mandate | Foster capital market growth | Foster capital market growth | Satisfy company capital needs |
| Regulator emphsis | Investor protection | Investor protection | Compliance |
| Pre-listing information quality | High | High | Medium |
| Pre-isting information disemination quality | High | High | Medium |
| IPO pricing mechanism | Demand-driven | Demand-driven | Value-driven |
| Time to establish intrinsic value | 1 month | 2 months | 1 month |
| Overreaction level | Medium | Low | High |
| Regulation post-listing | Weak | Weak | Strong |
| Self-regulation/Self-discipline | Weak | Weak | Strong |
| Delisting rules | None | None | Yes |
| Regulator resources required pre-listing | High | High | Medium |
| Regulator resources required post-listing | Low | Low | Medium |
| IPO Pricing Efficiency | TBD (need long-term performance) | Low | TBD (need long-term performance) |